# Fragmentation Considered Poisonous

*Amir Herzberg*[†] *and Haya Shulman*[‡]
*Dept. of Computer Science, Bar Ilan University*
[†]*amir.herzberg@gmail.com,* [‡]*haya.shulman@gmail.com*

## Abstract

We present practical *poisoning* and *name-server blocking* attacks on standard DNS resolvers, by *off-path, spoofing adversaries*. Our attacks exploit large DNS responses that cause IP fragmentation; such long responses are increasingly common, mainly due to the use of DNSSEC.

In common scenarios, where DNSSEC is partially or incorrectly deployed, our poisoning attacks allow 'complete' domain hijacking. When DNSSEC is fully deployed, attacker can force use of fake name server; we show exploits of this allowing off-path traffic analysis and covert channel. When using NSEC3 opt-out, attacker can also create fake subdomains, circumventing same origin restrictions. Our attacks circumvent resolver-side defenses, e.g., port randomisation, IP randomisation and query randomisation.

The (new) *name server (NS) blocking* attacks force resolver to use specific name server. This attack allows *Degradation of Service, traffic-analysis and covert channel*, and also facilitates DNS poisoning.

We validated the attacks using standard resolver software and standard DNS name servers and zones, e.g., org.

## 1 Introduction

The correctness and availability of information in the Domain Name System are crucial for the operation of the Internet. In particular, DNS poisoning is a significant threat to Internet security, and can be used for phishing, credentials stealing (e.g., XSS), sending spam and phishing emails, eavesdropping, and more. Due to its wide use and universal availability, the DNS is also abused in other ways for different goals, including Denial of Service, fast-flux, covert botnet communication, and more.

Most DNS (poisoning and other) attacks 'in the wild', as well as most research, considered an *off-path adversary* that is able to send spoofed packets (but not to intercept, modify or block packets). The most well known is Kaminsky's DNS poisoning attack [21], which was exceedingly effective against many resolvers at the time (2008). Kaminsky's attack, and most other known DNS poisoning attacks, allows the attacker to cause resolvers to provide incorrect (poisoned) responses to DNS queries of the clients, and thereby 'hijack' a domain name. We refer to this type of attack as *Domain-hijacking DNS poisoning*, to distinguish between it and two other variants of DNS poisoning, to which we also present off-path attacks; more below (and see Table 1).

The increased awareness to the risks of DNS poisoning, following the publication of Kaminsky's attack, was one of the factors helping to speed-up the (long-overdue) adoption of the DNSSEC standard [2–4]. DNSSEC uses digital signatures to prevent poisoning of the DNS responses, and is hence secure even against man-in-the-middle (MitM) adversaries. DNSSEC had a long and thorough design and evaluation process, and its security is based on extensive evaluation by many experts, as well as on results of formal analysis, e.g., by Bau and Mitchell [5]. (There are also alternate proposals for security against MitM, e.g., DNS Curve [7]).

However, deployment of DNSSEC is challenging and would take considerable time. Hence, as a more immediate response to Kaminsky's attack, resolvers were rapidly 'patched' to protect against it, mostly following RFC5452 [19]. The 'patches' include *source port randomisation, source and destination IP randomisation*, and *query randomisation: case toggling* ('0x20 encoding', [9]) and addition of a *random prefix* to queries [30].

These 'patches' increase the entropy of 'unpredictable' fields copied from DNS requests to DNS responses, and hence are trivially insecure against MitM attackers. However, these patches are considered 'best practice', since they are believed to provide sufficient security against off-path attackers (which is considered as sufficient defenses for many systems).

We show that, under common scenarios - which seem likely to become even more common in the future, an off-path attacker can efficiently circumvent all of these mechanisms. Namely, an off-path attacker can perform effective 'domain-hijacking' DNS poisoning attack, circumventing all the 'patches', and with even better efficiency than Kaminsky's poisoning technique. However, our domain-hijacking attack fails when DNSSEC is correctly and fully deployed. This attack works in scenarios where DNSSEC is partially or incorrectly deployed, such as permissive resolvers and islands of trust; currently, such scenarios are rather common.

We also present three other attacks, which apply even if DNSSEC is correctly and universally deployed:

**Subdomain Injection,** is a poisoning attack which causes resolvers to accept, cache and provide to clients a mapping for a non-existing (child) domain, of a DNSSEC-protected parent domain. As [5] observed, when the parent zone supports NSEC3 opt-out, the attacker can create fake (non-existing) sub-domains; this can lead to attacks on Same Origin Policy such as XSS, phishing and cookie stealing. We show how an off-path attacker can achieve the same effect.

**Name Server (NS) Hijacking,** is a poisoning attack which causes resolvers to cache and use incorrect name servers for a DNSSEC-protected domain, typically, pointing them to name servers belonging to the attacker. We show how this attack provides new, efficient off-path methods for traffic analysis, covert channels and Denial of Service; notice the attack and its applications are viable, even if DNSSEC is universally and correctly deployed, since the delegation NS and A resource records (RRs) are not signed.

**Name Server (NS) Blocking,** allows the attacker to force resolvers to query name servers of his choice, and stop using other name servers, by corrupting fragmented responses from those name servers. This is not a poisoning attack, since it does not involve a resolver accepting fake resource records. In fact, one application of this attack is to *facilitate poisoning*, by causing resolvers to use specific name servers, possibly known to be vulnerable; this provides an effective mechanism to deploy the attack of [32], which was, so far, considered impractical. Under certain situations, this attack can also be used for off-path Degradation of Service and traffic analysis.

We summarise all our attacks, with their requirements, in Table 1. The requirements are explained in Section 2; for now, it suffices to mention that they all reflect common situations in the current DNS, many of which are not expected to change, even if DNSSEC is fully, universally and correctly deployed. The main exception is attacks which require partial or incorrect DNSSEC deployment; however, not only is this requirement currently often satisfied, but it is also required only for the 'domain hijacking' attack.

In fact, ironically, the use of DNSSEC is often what provides necessary requirements for our attacks to work. Specifically, all of our attacks require *'Fragmentable zone'*, implying fragmented DNS responses; and three of the attacks require *'Poisonable zone'*, implying that the second fragment contains complete resource record(s), from the 'authority' and/or 'additional' sections. More details on the requirements are presented within.

DNSSEC requires long resource records (RRs) which results in long DNS responses. Long DNS responses (i.e., above 512 byte) require support of the EDNS extension mechanism, [35], and often fragmented when sent over UDP, since their size exceeds the path MTU. It is exactly this fragmentation that facilitates our attacks; e.g., we show that off-path attackers can often replace the second fragment of a packet, resulting in a seemingly-valid, yet fake, DNS response, or 'merely' causing corruption of the DNS response.

Fragmentation is known to be problematic or 'harmful', mainly due to the negative impact on performance; see the seminal paper of Kent and Mogul [23]. As a result, fragmentation is usually avoided, e.g., by use of path MTU discovery [28, 29], mainly for connection-based transport protocol (TCP). However, DNS traffic is usually sent over UDP; while several significant name servers, e.g., com, edu, send long responses over TCP, this may not be a good long-term solution, since the use of TCP results in significant overhead.

**Related Work.**

Our work builds upon previous disclosures of vulnerabilities due to the design or implementations of the fragmentation mechanism; we next mention few of the most relevant. Zalewski [37] suggested that it may be possible to spoof a non-first fragment of a (fragmented) TCP packet. However, using such non-first-fragment injections to TCP packets seems challenging. Furthermore, currently almost all TCP implementations use path MTU discovery [28,29] and avoid fragmentation.

Several vulnerabilities related to IP fragmentation, and specifically to predictable fragment identifiers (IP-ID) values, are covered in [14, 15]. Later, predictable IP-ID values were shown [12] to allow interception and injection of fragments, as well as dropping of fragmented packets. While for our purposes, a simple, randomised modification of fragments suffices, we modify the technique from [12] to improve the efficiency of the attack in some scenarios; details within.

| Attacks / Requirements | DNS Poisoning (Section 4) | | | Name Server Blocking |
|---|---|---|---|---|
| | Domain Hijacking | Subdomain Injection | NS Hijacking | |
| | Section 4.1 | Section 4.2 | Section 4.3 | Section 3.2 |
| IP-ID | √ | √ | √ | √ |
| ‘Fragmentable zone’ | √ | √ | √ | √ |
| ‘Poisonable zone’ | √ | √ | √ | |
| ‘Permissive or Island’ | √ | | | |
| NSEC3 opt-out | | √ | | |
| RFC 4697 | | | | √ |

Table 1: Our attacks and requirements.

**Attacker Capabilities.**

The required attacker capabilities include an arbitrary off-path, spoofing-only adversary, that controls a ‘puppet’, i.e., malicious (but sandboxed) script which can query the resolver; see Figure 1.

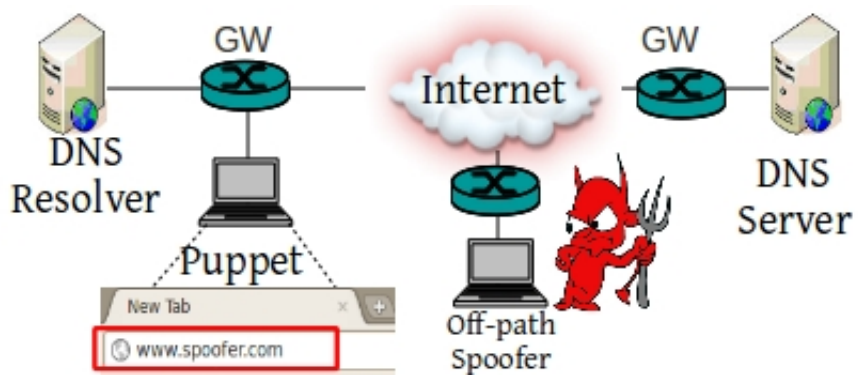

Figure 1: Simplified network topology and attacker capabilities. The spoofing attacker uses a puppet to invoke the query.

**Contributions.**

**Incremental DNSSEC Deployment is Vulnerable** We show that incremental deployment of DNSSEC is risky, and exposes resolvers to cache poisoning attacks. We present techniques which allow efficient and effective Kaminsky-style [22] cache poisoning attacks, using off-path spoofing adversaries. Our techniques (exploiting fragmentation) allow to circumvent the entropy (randomisation of query, source port and IP) in DNS responses, as those entropy fields remain in the first fragment. This exact technique of circumventing the entropy fields allows us to revive the Kaminsky attack, and to replace the authentic records in a DNS response with spoofed records.

**Subdomain Injection** We show that an off-path attacker can perform the attack that was believed to be feasible for MitM.

**Unsigned Delegation** We suggest attacks exploiting unsigned NS and A delegation records, breaching privacy and anonymity, and inflicting denial/degradation of service.

**Name Server (NS) Blocking** We introduce the name server blocking technique, which allows an attacker to force the resolver to stop using a particular name server, and eventually, to query a name server of attacker’s choice, e.g., a compromised name server, when resolvers strictly follow RFC 4697 [25].

We performed an experimental evaluation of our attacks against standard resolver software and issuing queries to real TLD name servers, such as org, and against the name server of the domain of our university, i.e., sec.cs.biu.ac.il.

**Organisation.**

In Section 2 we outline the assumptions which are required for our attacks and in Section 3.1 we provide the basic mechanisms for our attacks. Then, in Section 3.2, we show techniques for name server blocking, and in Section 4 we present the poisoning attacks. We then conclude and propose defenses in Section 5.

## 2 Attacks Requirements

In this section we describe the requirements of the different attacks that we introduce in this work. See Table 1 for the requirements of each attack. Below, we dedicate one subsection to the ‘Fragmentable zone’ and ‘Poisonable zone’ requirements, and another one to the ‘Permissive or Island’ requirement. We initiate with a brief description of three technical requirements: IP-ID, NSEC3 opt-out and RFC 4697.

### 2.1 Technical Requirements

The *IP-ID requirement* is that attackers have ‘reasonable’ probability of success in guessing the IP-ID in the responses from the name servers. In IPv4, the IP-ID field consists of only 16 bits; considering that fragment reassembly can usually hold a significant number of fragments for specific senders, typically at least 64, this implies a good success probability even if attacker just guesses the IP-ID values. Furthermore, many systems, and in particular most name servers, authoritative for major TLD zones, e.g., mil, use operating systems where IP-ID is generated sequentially, either globally (for all destinations) or with per-destination counters. In both cases, we can significantly improve the probability for a correct match. We optimised the IP-ID prediction, in the per-destination case, by adapting the technique of [12] to be used with resolver. Note that in IPv6, it is harder to predict the IP-ID, since it is 32 bits and is sent only in fragmented traffic; however, IPv6 specifically recommends the use of sequential IP-ID, and hence can be guessed with good chance of success. The bottom line is that the IP-ID requirement is almost always satisfied.

The *NSEC3 opt-out requirement* is that the zone uses NSEC3 DNSSEC record with opt-out option [26]. This allows attackers to create fake (non-existing) subdomains, and thereby facilitate XSS, phishing and cookie stealing attacks. Sub-domain injection attack was proposed in [5], that carried it out by a MitM attacker. In fact, [5], also suggested that the attack could be carried out by an off-path attacker, assuming that *only* the transaction ID in DNS packets is randomised. However, this assumption does not hold for most DNS resolvers, as they (at the very least) support source port randomisation. In Section 4.2 we show that such an attack can be effectively carried out by an off-path attacker, that does not intercept and inspect packets, and against patched DNS resolvers, i.e., supporting source port randomisation, IP randomisation and DNS query randomisation.

In spite of the publication of this potential abuse by MitM [5], NSEC3 opt-out is still widely used, and often even recommended, since it improves performance (esp. as long as DNSSEC is deployed only in small fraction of the domains).

The *RFC 4697 requirement* is that resolvers adhere to (one of) the recommendations in RFC 4697 [25, 36], specifically, that a resolver will refrain from sending queries to a name server after (few) failures, i.e., timeout queries, within a predefined time interval, and to *lame* name servers, i.e., those that provide DNS responses where (some of) the DNSSEC records, e.g., signatures, are missing or corrupted. We show in Section 3.2 that this allows an attacker to cause resolvers to stop using specific name server(s), facilitating poisoning attacks, as well as other attacks, including off-path traffic analysis and covert channel attacks. We validated this behaviour in popular and standard DNS resolvers, see Section 3.2.

## 2.2 Fragmentable Zone and Poisonable Zone Requirements

We now describe our two fragmentation requirements: ‘Fragmentable zone’ and ‘Poisonable zone’. The *‘Fragmentable zone’* requirement is necesssary to all of our attacks, and essentially amounts to the ability of the attacker to cause fragmentation of a response from a name server. As can be seen in Figure 2, long second fragments are rather common, and responses are fragmented from most top level domains (which deploy DNSSEC), e.g., in the gov top-level domain, which is the top-level domain with maximal adoption of DNSSEC so far.

Note that fragmentation can also occur even if the resolver does not deploy DNSSEC, e.g., it relays a query for a DNSKEY from the client, or due to other types of records which can be long, e.g., different TXT records, and/or when fragmentation occurs at relatively low packet length $l$, due to success in sending fake ICMP fragmentation-required alerts (see Section 2.2).

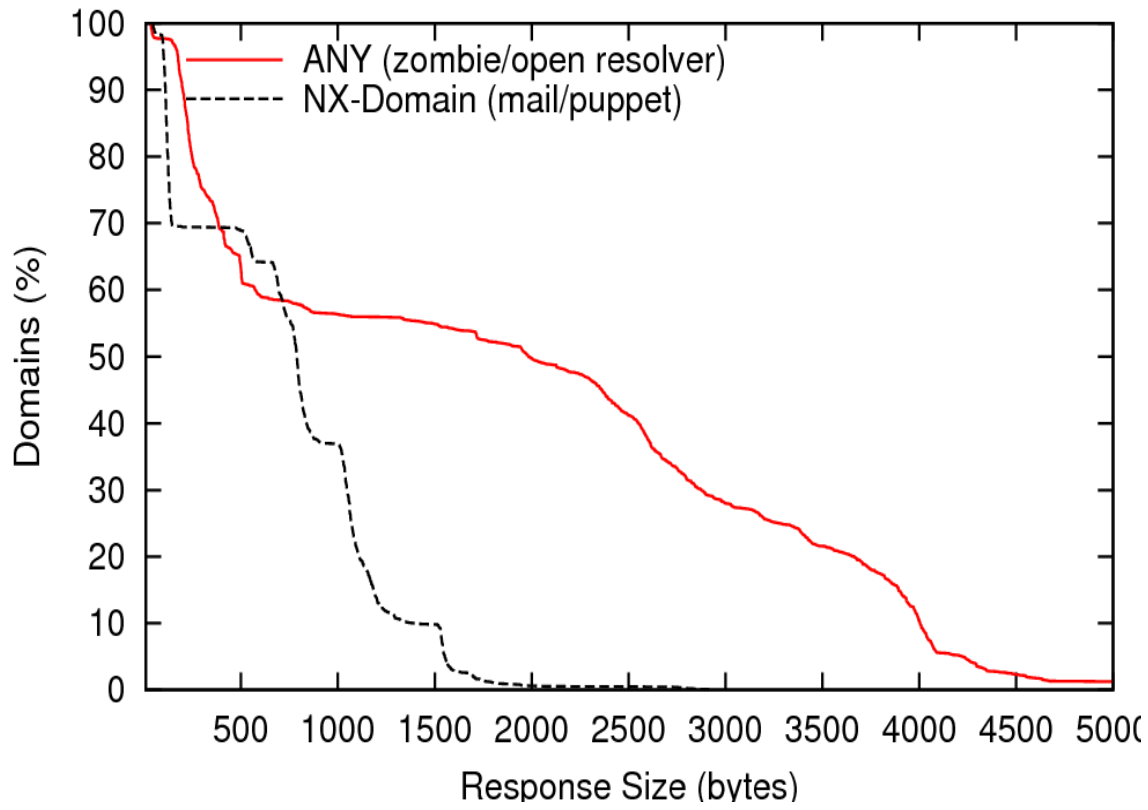

Figure 2: Length of ANY and NXDOMAIN responses of gov domains. Domains taken from [10]

The *‘Poisonable zone’* requirement is necessary for our poisoning attacks, and requires that at least one complete record is present in the second fragment so that the attacker can modify the second fragment and replace the authentic resource record with a spoofed one, which gets accepted and cached by the resolver. Details follow.

The ability to cause the required fragmentation depends both on the records in the zone file, as well as on the properties of the name server (including the route from the name server to the resolver). In particular, fragmentation typically depends on the smallest MTU enroute on the path from the name server to the resolver; packets larger than 1500 bytes will normally be fragmented. The attacker may also sometimes be able to ‘trick’ the name server into fragmenting (even) shorter packets before sending them. The attacker can inflict fragmentation by sending a spoofed ICMP 'fragmentation needed' packet, similar to attacks in [13]. Let $l$ denote the size of packets, above which attacker can cause fragmentation - the minimal MTU along the path, or the minimal length that the attacker can 'force' the name server to use by issuing fake ICMP fragmentation-needed alerts.

In principle, name servers could be configured to never allow fragmentation of responses, by sending response packets of length up to (some bound of) $l$ bytes with the DF (do not fragment) bit set, and send longer responses only via TCP, relying on TCP's path MTU discovery mechanism. However, the use of TCP for DNS requests and responses has significant performance penalty, in addition to the overhead and complexity of handling fragmentation-required ICMP alerts received due to sending packets with the DF set, which reach a network whose MTU is smaller than the packet size. Hence, we do not expect name servers to send packets with DF bit set (and indeed have not seen this behavior, e.g., com, edu).

Note that for the 'Fragmentable zone' requirement to hold, any fragmentation suffices, e.g., of 1 byte, and there is no requirement on the contents or length of the second fragments. Hence, we only require existence of a response whose length is greater than $l$.

A 'Poisonable zone' requirement is a stronger assumption, since it also implies that the attacker is able to include a (fake) resource record in the response, such that the response - and in particular the fake record - would pass validation at the resolver and get cached. This requires that the second fragment is predictable[1] (to allow attacker to avoid corrupting the checksum), and that it contains at least one modifiable record - typically, an NS record in the `authority` section, or an A ('glue') record in the `additional` section. The challenge here is mainly to find queries which will result in second fragments containing the necessary record(s) which the attacker will replace with its own.

CACHING AND TIME TO LIVE (TTL). The DNS resolvers will not issue queries at all, if there is a corresponding cached response. The TTL field of each DNS resource record indicates how long it may be cached by resolvers.

The majority of TTLs of DNS records range between one hour to one day, [20]. However, many records have very low or even zero TTLs, e.g., records of content distribution networks (CDNs). Furthermore, some queries, most notably for non-existent domain, always or usually would not be in cache. In fact issuing queries for non-existent domain is similar to Kaminsky style attack, and allows to initiate the attack as frequently as required since the attacker simply prepends a new random string to the query. We demonstrated attacks exploiting fragmented *non-existent domain*[2] or *no-data*[3], [1], DNSSEC-enabled responses. We also found that often DNSSEC public verification key (DNSKEY) records, which typically exceed MTU and get fragmented, have relatively short TTL e.g., 15 minutes in org domain; they also often indicate short expiration time in the signatures.

Hence, caching would usually not prevent the attack, and the expiration time of some record from the cache can be anticipated by the attacker; there may be some impact on length of attack and possibly on its communication costs too, but this would not make the attack infeasible. Usually, successful poisoning happens within reasonable time.

## 2.3 'Permissive or Island' requirement

The 'Permissive or Island' requirement is that DNSSEC validation is either not used or is not *correctly* used, and thus ignored by the resolver; 'Island' means that not *all* the zones from the root to the target zone deploy DNSSEC correctly, and 'Permissive' means that resolver does not fail even DNSSEC-enabled responses do not validate. In either of this cases, DNSSEC does not offer protection, although deployed.

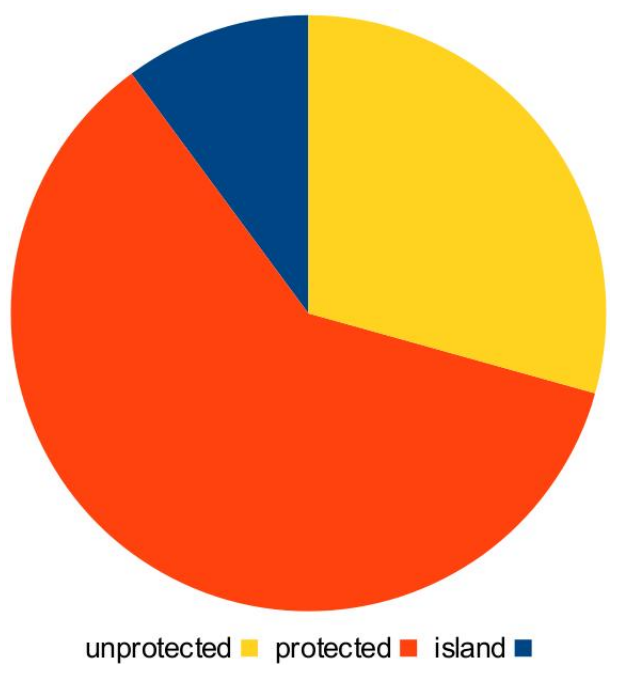

Figure 3: The DNSSEC deployment, in the list from [10] of 1500 subdomains of gov.

**Permissive Resolvers.** A permissive resolver is one that supports DNSSEC, however, ignores DNSSEC validation failures, e.g., if the signatures are missing or invalid; Unbound DNS resolver has an explicit 'permissive' mode to support such operation. Obviously, for such resolvers, DNSSEC does not provide added security; yet, there appears to be a significant number of such resolvers [8,16], apparently due to concerns about loss of connectivity due to interoperability and other problems upon enforcing DNSSEC. Such implementors deploy DNSSEC incorrectly or possibly via an 'incremental deployment', aiming to preserve DNS functionality with intermediate Internet devices, e.g., firewalls, and legacy resolvers which may discard DNSSEC enabled DNS responses or strip DNSSEC signatures. This approach apparently assumes that permissive use of DNSSEC can provide evidence on whether the network can deploy DNSSEC fully without problems or not, while not *harming* their security; unfortunately, this is not the case and such resolvers are open to poisoning.

[1]The (off-path) attacker can query the victim name server itself to select the query whose response it will poison.

[2]Non-existent domain DNS response contains an error bit in the DNS header, i.e., RCODE='name error', and indicates that the requested name does not exists in the zone file.

[3]No data DNS response does not contain an error, i.e., RCODE='no error', and it means that the requested name exists in the zone file but does not have the type requested in the DNS query.

**Island of Security.** When the parent zone returns an NSEC or NSEC3 (indicating that the child does not support DNSSEC), while the child does, the resolver cannot establish a 'chain-of-trust' to the target zone and thus cannot validate the DNSKEY of the zone. This holds for many second level domains, e.g., children of gov TLD (Figure 3), and even for some TLDs, e.g., mil. As a result, the resolver falls back to non-validating mode, and ignores the signatures in a DNS response.

## 3 Fragment Overwriting and NS-Blocking

In this section, we present the *Name Server blocking (NS-blocking) attack*, allowing attackers to cause resolvers to avoid querying a name server. This attack requires much weaker assumptions than the poisoning attacks, Section 4, and only assumes that the responses get fragmented. Both NS-blocking and the poisoning attacks, rely on *Fragment Overwriting*.

Fragment Overwriting, explained next, allows attackers to modify fragmented packets. In Section 3.2 we explain how fragment overwriting allows an attacker to cause standard-conforming resolvers to avoid a particular name server, i.e., to perform NS-blocking. Then, in Section 3.3, we discuss potential exploits of NS-blocking.

### 3.1 Fragment Overwriting

We now present a simple technique, allowing an attacker to replace (overwrite) second fragments of (fragmented) IP packets, in particular, of DNS responses sent to the resolver; this basic technique is applied in all of our attacks.

Suppose the original packet has payload $x$, and, after fragmentation, it is sent as two IP packets $y_1, y_2$. Then, the defragmentation process, running at the resolver[4] on inputs $< y_1, y_2 >$ reproduces $x$.

To overwrite the second fragment, the attacker sends a fake second fragment $y_2'$ so that it arrives at the defragmentation module *before* the authentic fragments $y_1, y_2$ of the DNS response. The defragmentation mechanism in the IP layer will cache $y_2'$, in anticipation of the rest of the packet. By default, an unmatched fragment is kept in the cache for 30 seconds or so, hence, 'planting' such fake fragments is easy; we now explain how the attacker can ensure a match with the original packet. Note that it is easy to adjust this technique for the (less common) case where fragments are sent in a reverse order: attacker removes the authentic second fragment $y_2$ from the reassembly buffer by sending an arbitrary $y_1'$ (whose validation fields match those of the $y_2$), and the rest is the same as above.

According to [18,31] the fragments of a datagram are associated with each other by their protocol number, the value in their IP-ID field, and by the source/destination IP address pair. Thus both the first authentic fragment $y_1$ and the second spoofed fragment $y_2'$ must have the same destination IP address (of the resolver that sent the query), the same source IP address (of the responding DNS server), the same protocol field (UDP) and the same fragment identifier (IP-ID). In addition, the spoofed second fragment should have the correct offset (as in the authentic second fragment). The fragment reassembly process, applied to the pair $< y_1, y_2' >$, returns either a failure or a different packet $x' \neq x$.

Matching most of these parameters is not difficult. In particular, path MTU changes infrequently, and can be found by attacker easily, e.g., by trace-route. In many scenarios, the resolver has a single, known IP address; zones typically have 6 IP addresses on average. The attacker may sometimes also have some knowledge on the likely name servers since the server selection algorithms of many resolvers can be predicted [36], e.g., based on latency, and may even use our attacks to disable some name servers; in the worst case, the attacker can launch the attack for each name server in parallel. In this work we show techniques which often allow the attacker to simply block the name servers of its choice.

It remains to ensure a match between the value of the fragment identifier (IP-ID) field in the fake fragment $y_2'$, and the IP-ID in the authentic fragment $y_1$ of the original response $x$. There are several possible options for IP-ID prediction, depending on the version of the IP protocol (IPv4 or IPv6), the method that the name server's OS uses to select IP-IDs, and on the receiver implementation.

In the most common case the communication is over

[4]Alternatively, defragmentation may happen and at intermediate device such as a firewall or NAT; there is no impact on the attack.

IPv4, where the IP-ID field is 16 bits; and the host performing the defragmentation process, i.e., the resolver or firewall, has cache of 64 (or more) fragments per particular <source IP, destination IP, protocol> combination; note that when more than 64 fragments (for the same tuple) arrive, the oldest fragment is evicted from cache and is replaced with the new one. Assuming that the attacker has no knowledge on the process according to which the IP-ID is incremented, and assuming the IP-ID in the packet and in the fake fragment are selected independently, it suffices for the attacker to send 64 spoofed second fragments, to ensure success probability of roughly $1/1000$ of replacing the second fragment of a packet in a *single* attempt. A much better probability of success (as we discuss next) can typically be achieved, however, even this attack can be sufficiently efficient for many scenarios.

Most systems select the IP-ID sequentially. Of these, many use a single counter for all destinations (globally-sequential), as in Windows and by default in FreeBSD. Other systems, e.g., Linux, use *per-destination* IP-ID counters. In both of these cases, the attacker can efficiently predict the IP-ID, achieving high probability of success (certainly compared to $1/1000$). In particular, for globally-incrementing IP-ID, which appears to be more widely used, e.g., mil, the attacker can simply learn the current value, and the rate of propagation, by querying the name server directly. The technique we use to achieve improved success probability in the case of per-destination IP-ID, is an improvement of the methods of [12]. These techniques ensure feasibility of the attack even for most implementations of IPv6, in spite of its use of 32-bits IP-ID field.

## 3.2 Name-Server Blocking

In this subsection, we present the NS-blocking attack, allowing an off-path attacker to dissuade resolvers from querying particular name server(s); this can have multiple goals, including denial/degradation of service, traffic analysis and more (see next subsection).

As indicated in Table 1, the NS-blocking attack has two main requirements: the ability of the attacker to generate a DNS query from the resolver, that will result in fragmented response from the 'victim' name server; and that the resolver follows a behaviour recommended in RFC 4697 [25, 36], namely, of avoiding a name server if there are two or more failures within some time interval (the time interval depends on the resolver implementation). We first describe how the attacker is able to block responses to a particular query, which does not rely on the behaviour recommended in [25].

To block responses to a query from a particular name server to the resolver, the attacker needs to send an arbitrary fake second fragment $y_2'$ with the anticipated IP-ID and other parameters, to match a legitimate response, as described above. The reassembly process using the legitimate first packet and the fake second fragment usually fails, and both fragments are silently dropped, by the UDP layer or by the DNS resolver itself, due to incorrect checksum. The resolver will resend the query after a timeout, but the timeout periods are known to the attacker, who can easily send appropriate fake fragments to cause loss of each of the responses, until the resolver gives up. The number of attempted retransmissions depends on the number of name servers that the zone uses; if the zone uses one name server, the resolver gives up in about 5 to 7 retransmissions, e.g., Bind9 after 7 timeouts, and Unbound 1.4.10 after 5 timeouts, if the zone uses more name servers then after one to two timeouts the resolver queries the next name server.

We show how to apply this technique, in order to perform NS-blocking, i.e., block a specific name server, instead of blocking just one particular packet. For this, we need the assumption that the resolver avoids querying unresponsive name servers, as per the recommendations in [25, 36]. We exploit the fact that when the target name server is not responsive, i.e., two or few queries times-out, the resolver does not send[5] any more queries to it.

The attack is illustrated in Figure 4. The idea is for the attacker to send a spoofed fragment, e.g., one byte long, which ruins the DNS response from the specific IP address. After repeating the attack few times (depending on the resolver software), e.g., twice in every 15 minute interval for Unbound, the resolver marks the name server (or rather the IP of the name server) as non-responsive and does not send queries to it for the specified interval of time which depends on the resolver implementation.

It is important to notice, that with most resolvers, NS-blocking is effective against a specific *name server IP*, and not limited to a specific domain. Namely, we can use NS-blocking to dissuade a resolver from using a specific name server, e.g., at IP address 38.103.2.1, using queries (with fragmented responses) to one domain isi-sns.info, and as a result the resolver will also avoid this name server for all the other domains which that server serves, e.g., paypal.com. This can be very useful, as name servers often host multiple domains, and some of these may use DNSSEC (and have fragmented responses), or be owned (or corrupted) by the attacker (who can put some other record with fragmented response), while others may only have short (unfragmented) responses, in addition, some domains are much less frequently queried, e.g., info, than popular domains, e.g., paypal.com, that use the same name server, allowing to match the IP-ID with little effort; e.g., [32] show that a typical domain name depends on 46 name servers on average which also serve other domains.

[5]Resolvers may send periodical probes, to detect when the target server becomes responsive, however, the period is so long that we can ignore it, e.g., for the Unbound name server, the period is 15 minutes. A similar behaviour of avoiding non-responsive name servers was observed by [36] in PowerDNS and WindowsDNS.

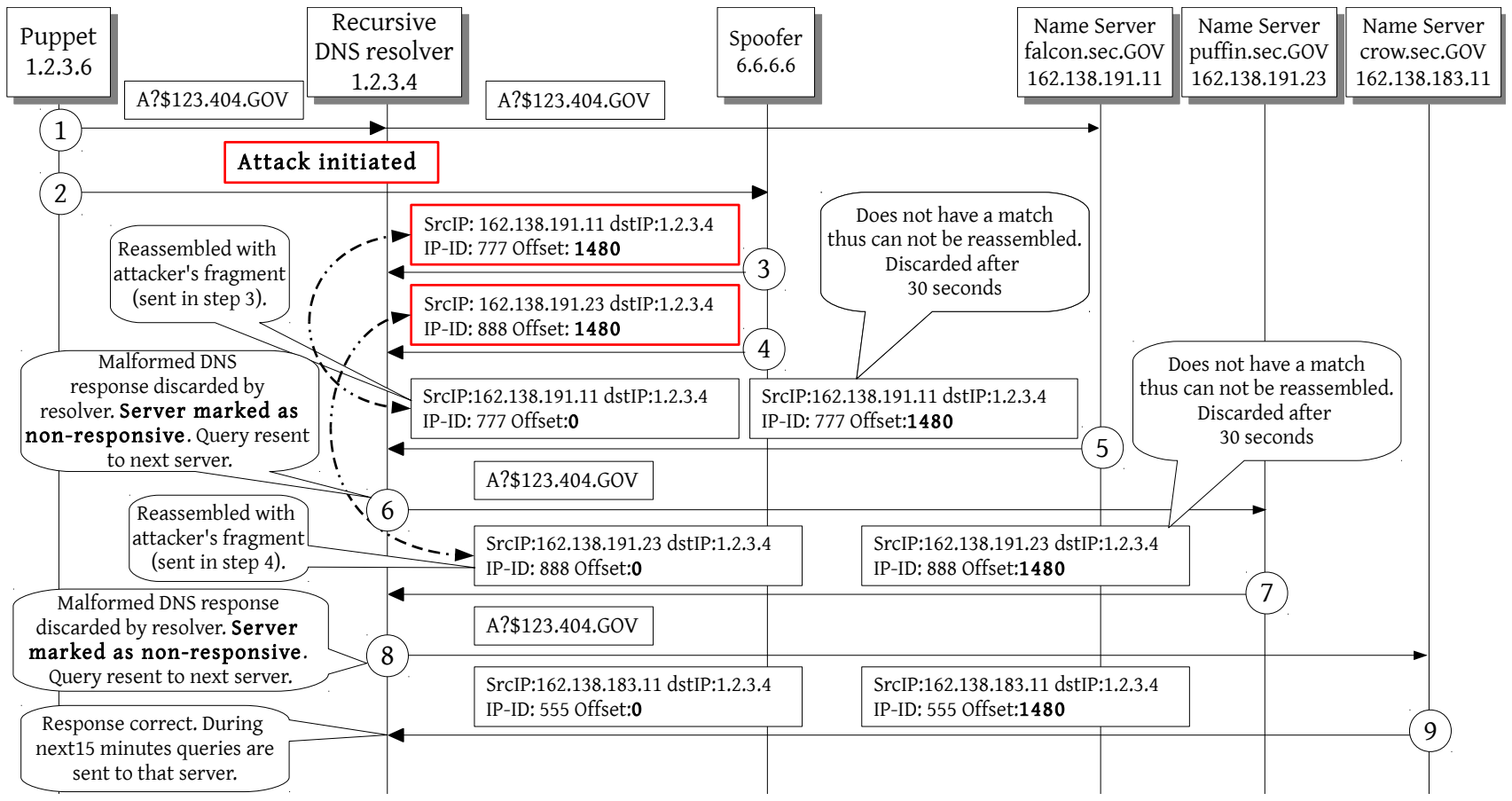

Figure 4: The Name-Server blocking attacks.

As illustrated in Figure 4, we performed the attack against a 404.gov domain, whose *non-existing domain* responses exceed 1500 bytes and thus get fragmented en-route. This phase, of forcing the resolver to use a specific IP, requires a puppet, i.e., a script confined in a browser, which issues DNS requests via the local caching DNS resolver, at IP 1.2.3.4 in Figure 4.

In steps 1 and 2 the puppet coordinates with the attacker and issues a DNS request for $123.404.gov (where $123 is a random prefix). In steps 3 and 4, the spoofer sends a forged second fragment, for all the possible name servers (i.e., a total of 2 spoofed fragments) except one which the attacker wants the resolver to use for its queries during the poisoning phase; the 404.gov domain has three name servers. This ensures that only one IP address will result in a valid response, and the other two result in a malformed DNS packets. The spoofed second fragment is incorrect, and contains a single arbitrary byte (in addition to headers). In step 5, the spoofed second fragment is reconstructed with the authentic first fragment resulting in a malformed DNS packet which leaves the fragments reassembly buffer. This malformed DNS response is then discarded by the resolver, and the IP of the name server is marked[6] as 'non-responsive'. When the authentic second fragment arrives, it does not have a match and is discarded after a timeout. As a result the resolver does not receive the response, and after a timeout it resends the DNS request to the next DNS server, step 6. The same procedure applies here, and the response is discarded. In step 9 a valid response arrives from IP 162.138.183.11. This way, by wrecking the responses from all name servers except one, we forced the resolver to direct all its queries for 404.gov domain to 162.138.183.11.

The IP-ID allocation algorithm does not have a significant impact on our attacks against such resolvers (e.g., Unbound), since 'misses', i.e., valid responses arriving to the resolver from some IP, do not prevent the attack; e.g., two failed (timed-out) queries suffice for Unbound to mark the server as non-responsive for 15 minute interval.

The Wireshark capture, in Figure 5, that was run on the resolver, demonstrates the experimental evalutation, i.e., the DNS packets entering/leaving the network card of the resolver. During the course of the experiment the puppet issued 6000 queries[7] to the resolver. The spoofer initiates the attack by sending three spoofed fragments to each IP address except 162.138.183.11. For simplicity, the capture presents only the packets exchanged between the resolver and the name server of 404.gov at 162.138.191.23 (by adjusting a corresponding filter in wireshark); the complete capture contains queries/responses from other name servers too. Packets numbered 18-20 are the forged fragments sent by the spoofer, with sequentially incrementing IP-IDs. Then puppet triggers a DNS request, packet 29. The response from the name server contains two fragments, packets 33 and 34. The first fragment is reassembled with spoofed fragment 18, resulting in a malformed packet which is discarded by the resolver.

The second fragment is discarded after a timeout. In packet 48 the resolver requests a public verification key of the 404.gov zone. The response contains three fragments 49-51; the first fragment is reconstructed with the spoofed fragment in packet 20, which also results in a malformed DNS response and is discarded. Note that this request, in packet 48, was sent at 19:28. Based on our tests it can be seen that when Unbound encounters a timeout twice for the same destination IP, it stops sending further packets to that destination for 15 minutes. Indeed, the next packet that is sent to that IP is packet number 6848, at time 19:43. The same scenario was observed with IP 162.138.191.11. The queries between 19:28 and 19:43 were sent only to 162.138.183.11, avoiding 162.138.191.11 and 162.183.191.23. Note that even if some of the responses (between packets 33 and 49) were valid and accepted by resolver, e.g., if they were not fragmented, it did not make a difference, and two timed-out responses in a 15 minute interval were sufficient for Unbound to stop querying those IP addresses; this fact shows that the success probability of the attack does not depend on the IP-ID selection mechanism.

[6]In reality the resolver marks the server as 'non-responsive' after two unsuccessful responses; this is handled by sending two spoofed fragments with consecutive IP-ID in IP headers.

[7]Note that our goal was to test the behaviour of the resolver, and to check the frequency of the queries to non-responsive servers; in real attack, once the IP-ID is known it is sufficient to issue two queries to mark the server as non-responsive.

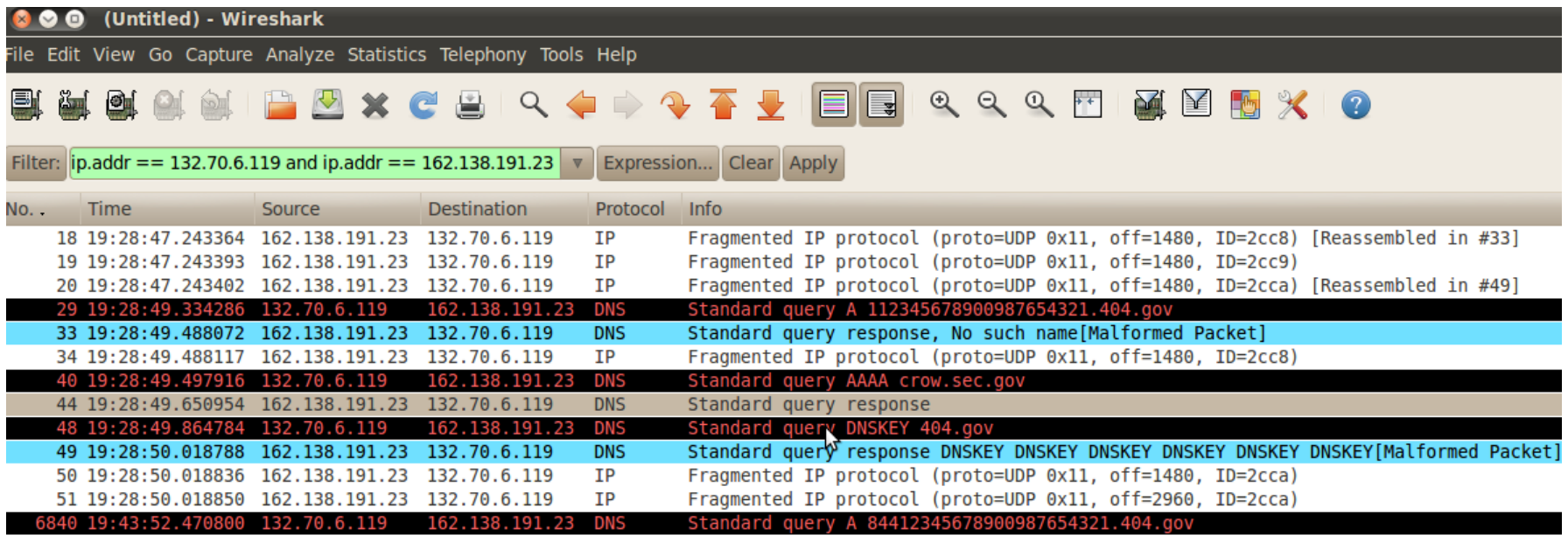

Figure 5: The wireshark capture of the attack, presenting only the packets exchanged between the name server 162.183.191.23 and the resolver. As can be observed, after two malformed responses the resolver refrains from sending further queries to that name server for 15 minutes. Fragmented packets are coloured in white, DNS requests in black, and reassembled DNS fragments in blue.

## 3.3 NS-Blocking: Applications

NS-blocking is rarely a goal by itself; more often, it can serve as a mechanism for other goals. We discuss three such goals: facilitation of DNS poisoning; degradation of service, and traffic analysis.

**Facilitate DNS poisoning** In [32], Ramasubramanian and Sirer conducted a survey showing that a typical domain name depends on 46 servers on average, and names belonging to countries depend on a few hundred servers. They note that compromising a server can lead to domain hijacks and postulate that it is possible to hijack 30% of the domains in Yahoo and DMOZ.org directories; DNS servers are known to have vulnerabilities [11, 24]. However, [32] did not suggest a technique which can be used to force a resolver to query a specific name server. NS-blocking can provide exactly the necessary mechanism.

Also note that NS-blocking can assist in other DNS cache poisoning attacks, including these in the next section of this paper, as it allows the attacker to reduce the number of servers that the resolver can query, possibly to only one.

**Off-path Degradation of Service** By blocking 'good' name servers, an attacker can cause resolvers to send their traffic to specific, 'bad' name servers. In particular, resolvers may resort to name servers with very high latency, causing unnecessary delays. Note that the zone administrators often deploy techniques to distribute the load between several physical servers sharing the same IP, e.g., using load balancing or Anycast [17]. Typically not all the name servers of a domain deploy such optimisations, e.g., 6 out of 13 root servers, [27]. Our technique allows the attacker to block those servers and to 'force' the resolver to query a name server which does not support such load balancing.

**Off-path Traffic Analysis and Covert Channel** Many names servers provide side-channels allowing an attacker to learn or estimate the rate of requests handled by the server. In particular, one simple and effective side-channel is the IP-ID used by the name server.

We next show how an off-path attacker can use this side channel, in conjunction with NS-blocking, to estimate (analyse) the rate of requests from some resolver $r$, to a particular domain foo.bar. NS-blocking can facilitate such off-path traffic analysis in several ways; as we later explain, this side-channel can even allow covert communication (between a bot using the resolver, and an attacker which is not controlling the name server).

First consider the case that one (or more) of the name servers of foo.bar, say ns.foo.bar, is using *globally-incrementing IP-IDs* (this is common). By using NS-blocking, we can direct all or most of the DNS requests from $r$ to ns.foo.bar; by periodically querying ns.foo.bar, the attacker can measure the rate of progress of the IP-ID (and hence of responses sent by ns.foo.bar). To further improve the measurement, the adversary may use NS-blocking to cause other major resolvers to avoid using ns.foo.bar.

This mechanism also allows *off-path covert channel*, between an agent, say a bot $b$, which can use the resolver $r$, and an off-path attacker $o$, which can make queries to the name server ns.foo.bar. The bot can, e.g., encode in-formation by signaling via the queries to ns.foo.bar (or possibly, signaling using distinct queries to several domains, each mapped to a specific, distinct name server). The attacker can communicate to the bot by signaling via loss of DNS responses.

The traffic analysis attack is applicable also if none of the name servers of foo.bar use globally-incrementing IP-ID, provided at least one of them, say ns.foo.bar, uses *per-destination incrementing IP-ID*. In this case, the attacker will also need the ability to use the resolver $r$, e.g., via a puppet (malware such as script, in a sandbox). Attacker will use the puppet to keep track of the IP-ID used by ns.foo.bar to send packets to $r$.

## 4 DNS Response Poisoning

The main idea behind our DNS cache poisoning attacks is to apply Second-fragment Overwriting technique to change the content of a DNS response by replacing authentic resource records (RRs) with spoofed A or NS RRs either for the existing domain or for a new subdomain. Specifically, the off-path attacker triggers a DNS request to some victim domain, e.g., using a puppet (malicious script confined in a browser), and then spoofs the second fragment. Depending on the section (of the DNS response) which the second fragment contains, i.e., `authority` or `additional`, the attacker replaces authentic records with his fake records[8].

The main difference between the attacks is related to the two intertwined issues: (1) frequency at which the attack can be repeated, and (2) the queries which the attacker can request. Both these issues depend on the freedom of the attacker over the (suitable) queries which it can request.

**NXDOMAIN or No-Data Responses.** To evade caching of the resolver the attacker can issue DNS requests for 'non-existing domain names', i.e., responses containing an RCODE with name error, or responses indicating that the domain name exists but with a different type, i.e., with 'no data no error' responses. Such responses exist in *every* domain. We found that domains that use NSEC3 (which is currently the majority of the domains) to indicate nxdomain (and no data) responses, to be most suitable for our attacks as those responses get fragmented in the `authority` section, such that the second fragment contains at least one complete RR (not including the EDNS RR). This technique is similar to Kamisky attack, as it allows the attacker to repeat the attack as frequently as required, by selecting a different random name (prepended to the real domain name) in each query. However, nxdomain responses may not always get fragmented, e.g., if the zone does not use NSEC3.

[8]The attacker can also replace the RRs in the `answer` section, e.g., if it can issue DNS requests for ANY type RR, and if the DNSSEC is either incorrectly, or not at all, deployed.

**'Existing Domain' Responses.** If the attacker triggers the attack with queries for existing domain names (with a non-zero TTL) that get fragmented, e.g., responses to DNSKEY, then it can trigger the attack only if the record is not in cache. If the attack fails the first time, it has to wait till the record expires from cache, i.e., the TTL reaches 0 or sinature expires.

Records with zero TTL, e.g., used for CDN networks, can be requested repeatedly, thus allowing to launch the attack as frequently as required, since they are not cached. Such DNS responses with (zero TTL) records in the `answer` section also contain (among others) NS and A records (in `authority` and `additional` sections respectively) with a cache-able TTL, which the attacker can replace. However, the zero TTL records may not exist in every domain, and in particular, may not exist in a domain which the attacker wishes to poison.

Thus the typically query choice of the attacker is between 'nxdomain/no-data' or 'exiting domain' (with a non-zero TTL).

**Poisoning the Cache.** Inserting spoofed records into the cache is not straight forward, and merely changing the records in a DNS response will not necessarily result in resolver accepting and (then) caching them. In particular, the forged DNS response must comply with the caching (and other) conditions which the resolvers impose on DNS responses: the forged packet $p'$ must be a valid DNS response; the 'poisonous' resource record $r'$ must be valid for the specific response and section of the response; and the resolver must cache $r'$.

The choice of forged DNS records, which we apply when modifying the DNS response, and semantics and values of each field in a DNS record, essentially follow the known rules for DNS cache poisoning and were investigated and studied (most notably) by Kaminsky [22] and Son and Shmatikov [33], and by Bau and Mitchell [5] for DNSSEC enabled DNS responses.

### 4.1 Domain Hijacking

Domain hijacking can be performed when the 'Permissive or Island' requirement is satisfied in tandem with 'Poisonable zone', i.e., either the resolver is permissive or zone is an island of security and there is at least one RR in the second fragment. In this case the attacker can replace the RR(s) in the authentic second fragment of a DNS response with (spoofed) NS or A RR(s) pointing at his name server; the TTL in those spoofed RRs has to match the TTL of the other RRs in that RRset.

Note that our domain hijacking techniques (below) trivially apply to (fragmented) DNS responses which are not protected with DNSSEC, we show such attack, exploiting queries for TXT RRs and the corresponding responses that get fragmented in full paper (the attack course is essentially similar to the attacks presented next).

We tested the domain hijacking attack in two scenarios (as elaborated above): (1) nxdomain/no-data and (2) existing domain. The 'nxdomain/no-data' (NSEC33) response is often fragmented in the `authority` section, and the `additional` section contains an[9] EDNS RR. This allows replacing the `authority` records in the second fragment with fake NS RRs; we show this attack in Section 4.1.1 replacing an NSEC3 record with a spoofed NS record in the `authority` section in response to a request for some non-exiting domain within sec.cs.biu.ac.il, i.e., the (DNSSEC-enabled) domain of the security group of the computer science department within our university.. The 'existing domain' response, e.g., DNSKEY or TXT, is also often fragmented. Such responses typically contain records in the `additional` section too, and allow changing the IP of name server with IP of the attacker. We show this attack in Section 4.1.2.

### 4.1.1 Injecting NS RR to NSEC3 Response

Typically, responses of type 'non-existing domain (nxdomain) name error' or 'no data no error', in domains that support NSEC3, are of size between 1700 to 2000 bytes on average, and when fragmented, at least one record from the authority section appears in the second fragment. This allows the attacker to replace the authentic NSEC3 or RRSIG RR(s) with a NS RR for a new name server; Figure 6. If the response does not contain any other NS RRs then the attacker can set an arbitrary high TTL, e.g., 6 days, to ensure that his RR stays in cache even when the authentic NS RRs for that domain expire. The attacker triggers a DNS request (via a puppet) and synchronises (steps 1 and 2, Figure 6). Then (step 3) the attacker sends a spoofed second fragment containing an NS RR for domain sec.cs.biu.ac.il. This spoofed fragment is combined with the authetic first fragment (step 3) and enters the cache; the authentic second fragment is discarded after a timeout (step 5). Note that the attacker can provide any arbitrary NS RR, in particular, one that is not in the same domain as the victim; in this attack we spoofed the response with name for a new NS RR, i.e., **ams**.sec.cs.biu.ac.il, in our domain, i.e., sec.cs.biu.ac.il, for testing purposes to observe that the subsequent queries of the resolver to domain sec.cs.biu.ac.il are sent to ams.sec.cs.biu.ac.il and responses get cached. To find the IP of the new NS the resolver initiates a request for the A RR, and receives and caches the IP supplied by the attacker (who controls that name server).

[9]This is not a requirement, and according to [1] an A RR can appear in the `additional` section too. In this case, the attacker simply replaces the A RR (instead of NS).

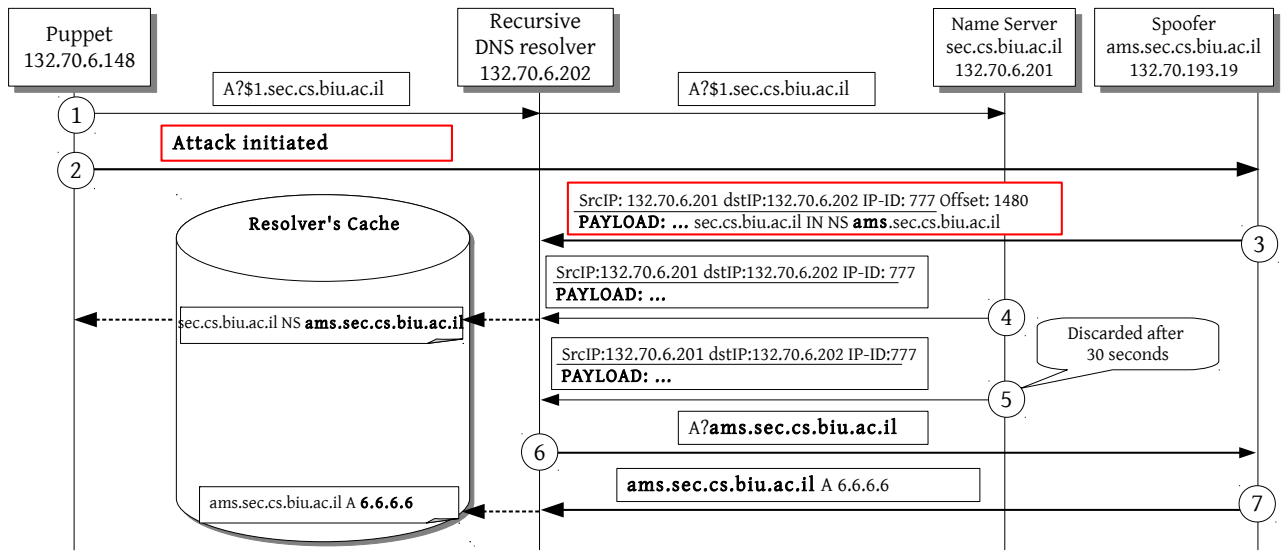

Figure 6: Poisoning an nxdomain (or no data) response, by replacing the NSEC3 RR with an NS RR.

The wireshark capture of the resulting poisoned DNS response is in Figure 7. The authentic fragment contained part of the RRSIG and two complete records, i.e., NSEC3 and a corresponding RRSIG. The spoofed fragment contained the authentic part of the RRSIG, spanning the first and second fragments, and two fake NS records which replaced the authentic NSEC3 and RRSIG. Note that since the RRSIG (as well as NSEC3) are much larger than NS RRs, the attacker has to pad the packet (with zeros) to the required length; the checksum is adjusted in the padded area after the EDNS RR. For a comparison, see the authentic DNS response in Figure 8.

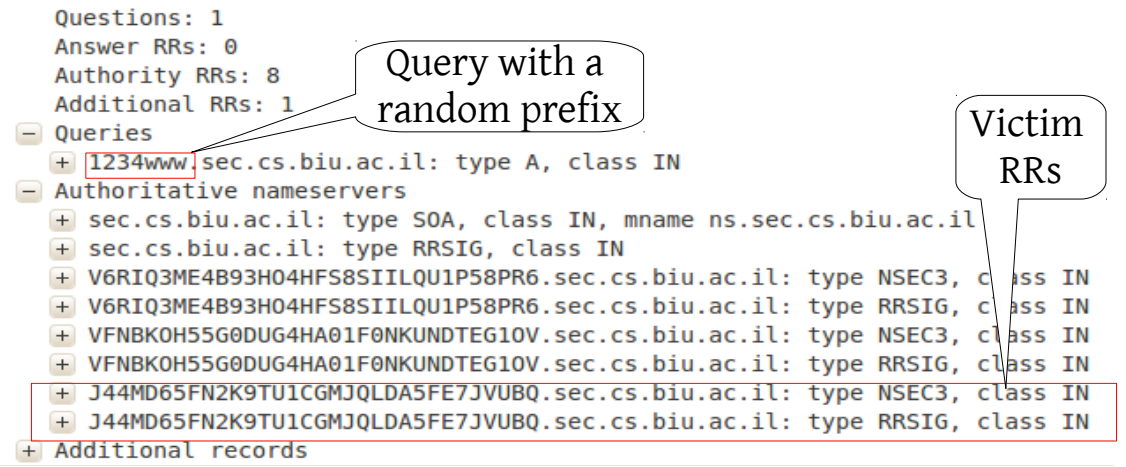

Figure 8: An authentic nxdomain response for domain sec.cs.biu.ac.il.

### 4.1.2 Injecting A RR to DNSKEY Response

When the second fragment contains at least one complete record (excluding the EDNS RR) in the `additional` section, the attacker can replace the IP address in the fragment with a spoofed IP. In this attack we spoof the IP for the name server of org domain, in a DNS response for a DNSKEY of org domain.

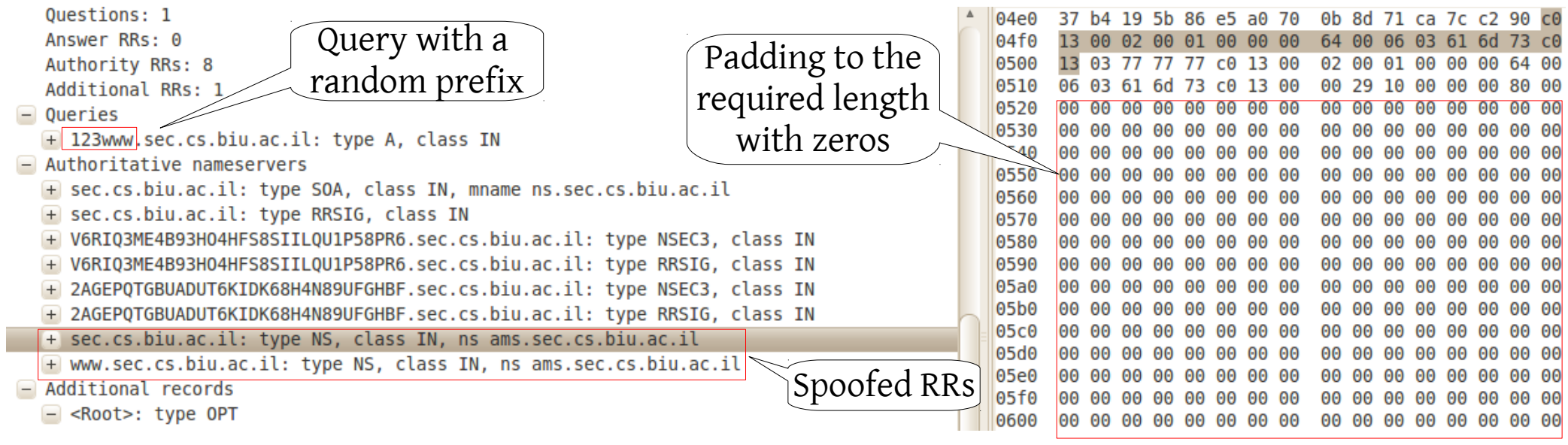

Figure 7: Poisoning an nxdomain (or no data) response for domain sec.cs.biu.ac.il, by replacing the NSEC3 RR with an NS RR.

In Figure 9 the resolver issues a DNS request for the DNSKEY of org; this is an indirect way to trigger a query, i.e., the resolver asks for the DNSKEY of some domain automatically, when the DNSKEY expires from cache, or when it needs to validate records for that domain, e.g., to be able to validate an A record or a non-existing domain (NSEC3) record; an attacker may also be able to cause a resolver, which does not support DNSSEC, to issue such a query, by sending an appropriate request to the resolver. This query type is useful if the response to an nxdomain query is not fragmented.

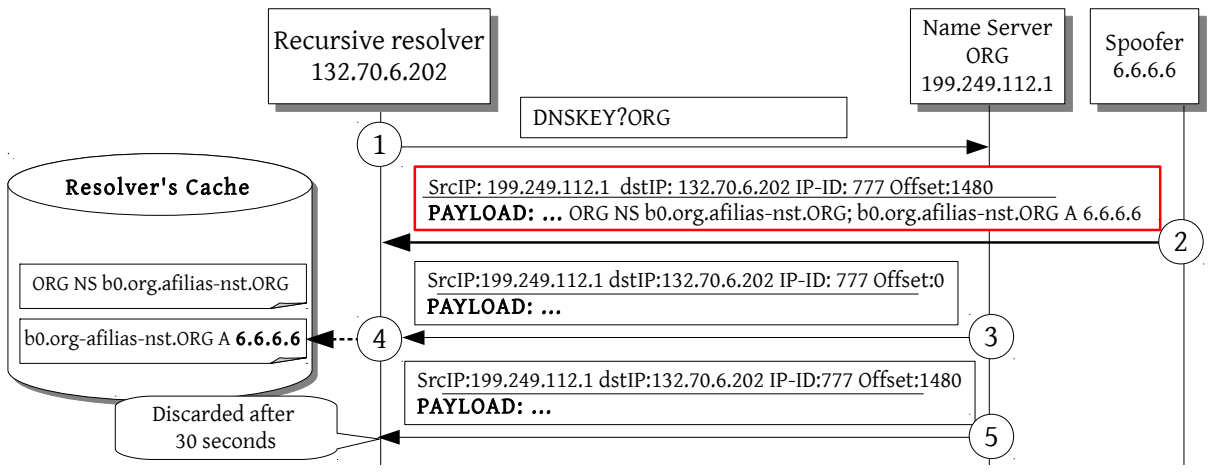

Figure 9: DNS poisoning attack of name server IP of org domain. The resolver issues a query for a DNSKEY (e.g., when it expires from cache, 15 minutes for org), and the spoofer sends a poisoned second fragment containing the forged entries of org. The query for DNSKEY of org can also be triggered indirectly by issuing a query for non-existing (or for some other) domain within org.

The annotated screen caption of the attack in Figure 9 is illustrated in Figure 10, presents the outcome of the attack. The first line (122) contains the 'forged second fragment'; this fragment is kept in the defragmentation cache of the resolver, waiting for a matching first fragment (i.e., with the same set of `(source IP=199.249.112.1, dest IP=132.70.6.202, fragment ID=7c6e, protocol=UDP)`). In the next line (133), the resolver sends the DNS query to the name server.

Next line (134) is the first fragment of authentic response to the query, sent by the name server of org (at IP 199.249.112.1). This response matches the fake second fragment already in the defragmentation cache, hence it appears as a complete DNS response packet. The contents of this packet are described in the lower panes; in particular, see the two forged resource records in the additional section, which contain incorrect (adversarial) IP addresses for two of the name servers of the org domain.

Finally, notice that the authentic second fragment, received in line 135, has no matching first fragment (since the one received was already reassembled with the spoofed second fragment). Hence, it is entered into the defragmentation cache, where it is likely to remain until discarded (typically, after maximal lifetime of about 30 seconds).

## 4.2 Subdomain Injection

The delegation records, NS (name server) and A (IP address), located in `authority` and `additional` sections, are not signed, [3]. This allows the attacker to change the IP address (in the `additional` section) of the name server of some victim domain, to its own address, or to add a new name server (in `authority` section) for the victim domain. Such NS and A records are usually cached and used, for queries to the specified domain; see [33]. At this point, the attacker managed to cause queries for the victim subdomain to be sent to a machine controlled by the attacker.

This vulnerability, of redirecting DNSSEC enabled DNS requests to malicious server by a man-in-the-middle, was pointed out by Bernstein [6], yet without a specific application for such an exploit. Bau and Mitchell, [5], refute Bernstein's claim of this being a vulnerability, by proving that it does not enable a man-in-the-middle attacker any additional capabilities, and conclude that it does not pose a significant threat. Indeed, not signing the delegation records does not break the design requirements defined for DNSSEC, [3]. However, it exposes to NS Hijacking (leading to a range of other attacks), Section 4.3, and to subdomain injection (as we discuss next).

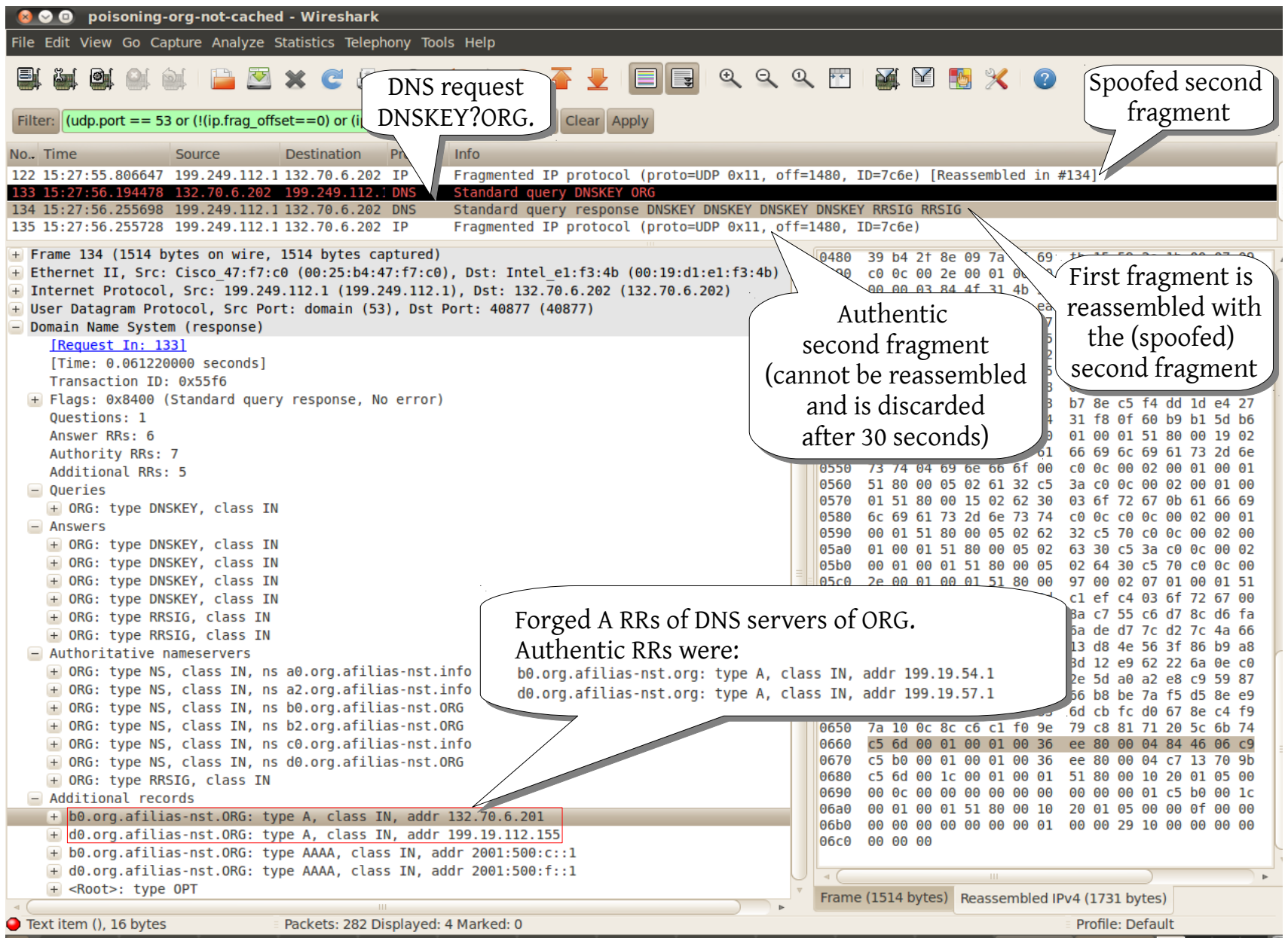

Figure 10: A Wireshark capture presenting the DNS packets in a DNS cache poisoning attack exploiting fragmented DNS responses by spoofing the second fragment of the response to the DNSKEY query; messages exhange diagram corresponding to this caption is in Figure 9.

Resource records in the `answer` section of DNSSEC-enabled domains are signed, hence, if the resolver performs strict validation of DNSSEC, it should not accept unsigned records in the `answer` section. However since the delegation records in the `authority` and `additional` sections are not signed, the attacker can create a new subdomain, e.g., secure.bank.com under bank.com, by creating an NS record (in the `authority` section) for the new subdomain secure.bank.com mapping it to the name server which the attacker controls. This attack is only applicable to DNSSEC protected domains that use NSEC3 opt-out: the attacker may be able to use legitimate NSEC3 responses, e.g., for non-existing sub-domain, or other type responses, e.g., DNSKEY, and turn them into what appears to be a legitimate referral to an unprotected subdomain; this can be used for phishing attacks. Furthermore, as [5] show, domains which deploy NSEC3 opt-out records are vulnerable to browser cookie theft by a man-in-the-middle attacker, as they allow the attacker to insert insecure delegations into DNS responses. This is in contrast to the recommendations in [26] which suggest that NSEC3 opt-out does not pose a significant threat. Bau and Mitchell performed the attack with a man-in-the-middle attacker, using our techniques an *off-path* spoofing attacker can perpetrate such attacks, see Figure 7.

The attack is similar to the attack in Section 4.1.1, thus to save space, we performed the addition of a new subdomain as part of the name server hijacking attack, see Figure 7. The record: www.sec.cs.biu.ac.il is a new (non-existing) subdomain under sec.cs.biu.ac.il and it points at the name server ams.sec.cs.biu.ac.il controlled by the attacker.

## 4.3 NS Hijacking

Using ‘server blocking’ the attacker can ‘fix’ the target name server. If the attacker compromised that server then the attack is very damaging. Server fixing can be useful for other attacks too, e.g., to degrade efficiency (if the target server is the slowest), for traffic analysis, e.g., if the attacker has man-in-the-middle capabilities but only on the path to that ‘fixed’ server, but not to other servers for that domain.

Server fixing in tandem with DNS poisoning can allow the attacker to force the resolver to use a malicious name server which the attacker controls, we call this ‘NS hijacking’. This attack is most relevant when DNSSEC is properly deployed. If the DNSSEC is not deployed correctly, then the attacker can simply hijack the domain. The attack is combined of two phases: (1) poisoning the A (or respectively NS) record in the DNS response (by changing the authentic IP to the IP controlled by the attacker), then (2) applying server blocking by ruining responses from all other name servers so that the resolver marks those authentic servers as non-responsive.

Note that phase (2), i.e., server blocking, is not es-sential and the poisoning attack by itself implies NS hijacking. This is due to the fact that the TTL of the poisoned RR is higher than the TTL of the records cached from previous responses, therefore, once those authentic records expire from cache, the resolver will not request them and will use the poisoned cached NS RR.

As a result of this attack the resolver will only query the server of the attacker (as it is the only one that responds). However, the attacker cannot produce valid signatures for the records that it returns, and therefore it responds to resolver's queries with records that are not protected with DNSSEC. This attack has the 'cache-or-crash' effect, i.e., the resolver will either cache those responses, or will timeout and not be able to provide responses (since this is the only name server that the resolver has for the victim domain). The response depends on the specific resolver in question, e.g., Unbound 1.4.1 in permissive mode caches such responses, while Bind9 times-out and does not.

## 5 Conclusions and Defenses

We showed how an off-path attacker can efficiently exploit fragmented DNS responses to poison DNS caches. Most DNS responses are short, and hence not fragmented; our attacks exploit DNSSEC records, which often result in fragmented responses.

We also showed that fragmented responses can be exploited by off-path attackers to force the resolvers to query name servers of attacker's choice.

The attacks are effective against valid implementation of the DNS and IP specifications; furthermore, we have confirmed effectiveness against several domains, using real network scenarios and common resolvers (Unbound 1.4.1 and Bind9).

We want to caution against drawing the conclusion that DNSSEC should not be used. In fact, the best defense is to apply DNSSEC correctly in *all* resolvers and domains (without using NSEC3 opt-out); this will certainly prevent many of our poisoning attacks, and even defend against more powerful Man-in-the-Middle adversaries. However, incremental DNSSEC deployment is vulnerable to our cache poisoning attacks, and complete adoption of DNSSEC may take considerable time, since it depends on adoption by both name server and resolver. Furthermore, this will not prevent the server blocking attacks. Hence, we also discuss some other defenses, which can be utilised by only one of the parties (unilaterally), and can also prevent the DNS response blocking attacks.

The vulnerability which allowed us to launch the poisoning attacks against recursive resolvers, is due to the fact that the resolver advertises a large EDNS buffer, which is usually larger than the MTU, e.g., 1500 bytes. Although support of large DNS responses is critical to facilitate DNSSEC enabled DNS responses, or public-key certificates [34], such long responses can be (temporarily) sent over TCP, using path MTU discovery and avoiding fragmentation - unfortunately, at significant performance costs. Another countermeasure, possible at resolver, name server or even at intermediate gateway (firewall), is to set a maximal EDNS buffer value to at most 1500, or even less, to avoid fragmentation. In fact, resolvers may implement a simple protocol similar to path MTU discovery, then set the value of the EDNS buffer accordingly to the minimal MTU en route between the resolver and the DNS server. When sending responses, the name server should also set the DF bit in the IP header to 1, i.e., do not fragment.

Another short-term defense, which administrators of resolvers can apply, is to reduce the maximal number of fragments cached; e.g., currently 64 by default in Linux (per (source IP , dest IP , protocol) triplet). Of course, reduction in this parameter may also increase packet loss.

Yet another possible defense for name servers, is to always add a *random RR* to any packet over certain size (i.e., which may be fragmented). A simple type A resource record, containing random IP address for some fictitious domain name, would suffice. The TTL of such an RR should be set to zero to prevent the resolver from caching that record. This would prevent the attacker from being able to predict and (correctly) adjust the checksum value. If there are multiple vulnerable fragments, such a random RR should appear in each fragment.

Finally, we suggest to be careful when deploying the proposal in [25, 36] (for server selection) which recommends to avoid querying non-responsive servers. Resolvers that do not conform to that recommendation, e.g., Bind9, are not vulnerable to our server-pinning attacks.